\documentclass[preprint,prb,amsmath,amssymb,amsfonts,superscriptaddress,floatfix,showpacs]{revtex4}

\usepackage{graphicx}
\usepackage{dcolumn}
\usepackage{subfigure}

\begin{document}

\title{The Delicate Electronic and Magnetic Structure \\
    of the LaOFePn System (Pn = pnictogen)}
\author{S.~Leb\`egue$^*$}
\affiliation{
Laboratoire de Cristallographie et de Mod\'elisation des Mat\'eriaux Min\'eraux et Biologiques, 
UMR 7036, CNRS-Universit\'e Henri Poincar\'e, B.P. 239, F-54506 Vandoeuvre-l\`es-Nancy, France
}
\author{Z. P. Yin\footnote{These authors contributed equally to this work.}}
\affiliation{
Department of Physics, University of California Davis, 
  Davis, CA 95616
  }
\author{W. E. Pickett}
\affiliation{
Department of Physics, University of California Davis, 
  Davis, CA 95616
  }

\date{\today}
\begin{abstract}
The occurrence of high temperature superconductivity, and the competition with magnetism,
in stoichiometric and doped LaOFeAs and isostructural iron-oxypnictides
is raising many fundamental questions about the 
electronic structure and magnetic interactions in this class of materials.  There are now
sufficient experimental data that it may be possible to identify the important issues
whose resolution will lead to the understanding of this system.  In this paper we address
a number of the important issues.  One important characteristic is the Fe-As distance
(or more abstractly the pnictogen (Pn) height $z$(Pn)); we present results for the effect
of $z$(Pn) on the electronic structure, energetics, and Fe magnetic moment.  We also study
LaOFeAs under pressure, and investigate the effects of both electron and hole doping within
the virtual crystal approximation.  The electric field gradients for all atoms in the 
LaOFeAs compound are presented (undoped and doped) and compared with available
data.  The observed $(\pi,\pi,\pi)$ magnetic
order is studied and compared with the computationally simpler
$(\pi,\pi,0)$ order which is probably a very good model in most respects.  We investigate
the crucial role of the pnictogen atom in this class, and predict the structures and properties
of the N and Sb counterparts that have not yet been reported experimentally.  At a certain
volume a gap opens at the Fermi level in LaOFeN, separating bonding from antibonding bands
and suggesting directions for a better simple understanding of the seemingly intricate
electronic structure of this system.  Finally, we address briefly on the possible effects of
post-lanthanum rare earths, which have been observed to enhance the superconducting
critical temperature substantially.
\end{abstract}
\maketitle

\section{Background and Motivation}
Isostructural and isovalent LaOFeP and LaOFeAs are layered conductors, 
the first being superconducting
at T$_c$=2.5 K\cite{Kamihara2006} while the second becomes antiferromagnetically ordered
at T$_N$ $\approx$ 140 K\cite{Cruz:arXiv0804.0795, Nomura:arXiv0804.3569} and is not superconducting.   
The discovery of superconductivity at 26 K in carrier-doped LaOFeAs\cite{Kamihara2008},
followed by rapid improvement now up to T$_c$=55 K\cite{Ren:arXiv0804.2053} in this class, makes
these superconductors second only to the cuprates in critical
temperature.  Several dozen preprints appeared within the two months
after the original publication, and many hundred since, making this the most active field of  
 new materials study in recent years (since the discovery in MgB$_2$,
at least).  

A host of models and ideas about the ``new physics'' that must be operating
in this class of compounds is appearing, pointing out the need to 
establish a clear underpinning of the basic electronic (and magnetic)
structure of the system.  The materials are strongly layered, 
quasi-two-dimensional in their electronic structure, by consensus.
The electronic structure of LaOFeP was described by Leb\`egue,\cite{Lebegue2007}
with the electronic structure and its neighboring magnetic instabilities
of LaOFeAs being provided by Singh and Du\cite{Singh:arXiv0803.0429}. Several illuminating
papers have appeared since, outlining various aspects of the electronic
and magnetic structure of LaOFeAs.

The extant electronic structure work has provided a great deal of necessary
information, but still leaves many questions unanswered, and indeed
some important questions are unaddressed so far.  In this paper we address
some of these questions more specifically. Stoichiometric LaOFeAs is 
AFM; then $\sim$0.05 carriers/Fe doping of either sign destroys magnetic order
and impressive superconductivity arises, with $T_c$ seemingly depending
little on the carrier concentration.  Another question is: with the
nonmagnetic electronic structure of LaOFeP and LaOFeAs being so similar, why is the 
former superconducting while the latter is (antiferro)magnetic?  Surely
this difference must be understood and built into bare-bones models, or
else such models risk explaining nothing, or explaining anything. 
Another question is the effect of the structure.  Unusual sensitivity to
the As height $z$(As) has been noted\cite{Yin:arXiv0804.3355}; 
T$_c$ is reported to increase
with applied pressure\cite{Takahashi2008,Lu-NJP-2008} 
 (reduction in volume) for low values of doping (up to $x=0.11$ in LaO$_{1-x}$F$_{x}$FeAs, which is reported as the amount of F
  for optimal doping);
there are increases in
T$_c$ due to replacement of La with other rare earth ions, and the
variation in size of the rare earth is often a dominant factor in the
observed trends in their compounds.  Very important also is the
magnetism in these materials, as magnetism is a central feature in
the cuprate superconductors and in correlated electron superconductors.
Another important question is: what can be expected if other pnictide
atoms can be incorporated into this system: Sb (or even Bi) on the large
atom side, or N on the small atom end.
In this paper we address these questions.

\section{\label{sec:two} Crystal Structure}

\begin{figure}[h]
\includegraphics*[angle=0,width=0.48\textwidth]{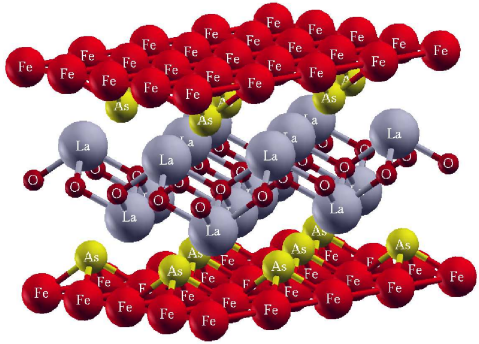}
\caption{(Color online) The crystal structure of LaOFeAs,
 showing the alternating layers of LaO and FeAs.
\label{fig:structure}
}
\end{figure}

The members of the family of the new Fe-based superconductors crystallize in the ZrCuSiAs type structure\cite{ZIMMER1995, Quebe2000}
(space group  P4/nmm, Z = 2). For instance, LaOFeAs is made of alternating LaO and FeAs layers, as presented in Fig. \ref{fig:structure}.
 The Fe and O atoms lie in planes, while the As and La atoms are distributed on each side of these planes following a chessboard pattern.
 The crystal structure is fully described by the a and c lattice parameters, together with the internal coordinates
  of La and As. Experimentally, $a = 4.03533$ \AA~and $c = 8.74090$ \AA, while z(La) = 0.14154 and z(As) = 0.6512.
 \begin{figure}[h]
\includegraphics*[angle=0,width=0.28\textwidth]{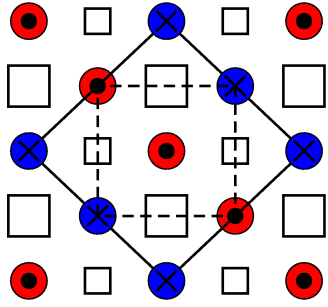}
\caption{(Color online) 
The Q$_M$ magnetic structure of the Fe-As substructure of LaOFeAs, 
showing alternating chains of Fe spin up (red circles
 with black dots) and Fe spin down (blue circles with black crosses). The As atoms above (below)
  the Fe plane are represented as large (small) squares. The $\sqrt{2}a \times \sqrt{2}a \times c$
   cell is represented in full lines, while the $a \times a \times c$ cell is in dashed lines.
\label{fig:afm}
}
\end{figure}
 However to describe correctly the antiferromagnetic structure, 
a $\sqrt{2}a \times \sqrt{2}a \times c$ cell must be used, with four Fe atoms
  per cell, as shown in full lines in Fig. \ref{fig:afm}. We will refer to this 
antiferromagnetic order as the Q$_M$ AFM order,
  or equivalently as ($\pi , \pi , 0$), while the Q$_0$ AFM order
  corresponds to an antiferromagnetic order of the original cell (dashed lines in Fig. \ref{fig:afm}) with two Fe atoms.
  Also, FM will refer to a ferromagnetic arrangement of the spins, while NM means non-magnetic.

\section{Calculation Method}
To calculate the relevant quantities, we have used density functional
theory (DFT) \cite{Hohenberg,Sham}, as implemented in 
three different electronic structure codes.
The full potential local orbital (FPLO) code\cite{fplo1, fplo2} 
was mainly used, 
while we double checked some of the calculations with Wien2k code \cite{wien2k}.
For most of the FPLO and LAPW calculations, the Perdew and Wang 1992 (PW92)\cite{PW92}
exchange-correlation (XC) functional was used, but the effect of XC functional was checked 
using also LSDA(PZ)\cite{PZ}, the PBE (Perdew {\it et al.} 1996)\cite{PBE}, 
and another GGA (Perdew {\it et al.} 1992)\cite{GGA-Perdew92}
 XC functionals. 
At each constant volume, the crystal 
structure was fully relaxed,
i.e., $c/a$, $z$(La) and $z$(Pn) were relaxed, where Pn is the pnictogen atom. 
The errors were estimated to be 
within 0.5$\%$ for c/a, and 1.0$\%$ for z(La) and z(Pn). 
The relaxation was performed in the Q$_M$ AFM structure,
with 132 irreducible k points in the BZ. We double checked the total energy 
with a finer mesh with 320 irreducible
k points in the BZ, and the difference is very small.
 After relaxation, 
all calculations were performed using dense meshes, with 320, 1027, and 637 
irreducible k points in the BZ of the
Q$_M$ AFM, Q$_0$ AFM and NM structure, respectively. In the Q$_M$ AFM structure, 
we used 464 irreducible k points in the BZ to double check the result, 
without any noticeable difference in the DOS nor band structure.
As for the results presented in Sect. \ref{sect-pipipi}, we used the PAW (projector augmented waves)
method \cite{Bloechl} as implemented in the code VASP (Vienna Ab-initio Simulation Package)\cite{vasp2,vasp}.
The Perdew Burke Ernzerhof\cite{PBE} variant of the generalized gradient approximation (GGA)
 was used for the exchange-correlation potential.
A cut-off of $600$ eV
 was used for the plane-wave expansion of the wave function to converge the relevant quantities.
 For Brillouin zone
  integrations, a mesh of $9 \times 9 \times 7 $ $k$-points\cite{Monkhorst} was used
within the modified tetrahedron method\cite{Bloechl94a}. This mesh was decreased to $9 \times 9 \times 3 $
 for the cell doubled along the $c$ axis.

\section{Study of LaOFeAs in the tetragonal structure}

LaOFeAs has a tetragonal structure (as described in Sect. \ref{sec:two}) at room temperature\cite{Kamihara2008}. 
Although it undergoes
a structural phase transition at lower temperature\cite{Cruz:arXiv0804.0795, Nomura:arXiv0804.3569}
 (see Section \ref{sect-pipipi} ), the doped (and superconducting)
  material LaO$_{1-x}$F$_{x}$FeAs remains in this structure down to low temperature,
   so the study of LaOFeAs in the high symmetry structure is a necessary step
    towards the understanding of the electronic structure of the whole family of compounds.

\subsection{Influence of XC functionals and codes on the electronic structure of LaOFeAs}

First, we studied the electronic structure of LaOFeAs in the experimental (tetragonal) crystal structure 
 for different magnetic states (Q$_M$ AFM, Q$_0$ AFM, FM and NM)
 using two different codes (FPLO$7$ and Wien$2$K) and different exchange-correlation functionals.
 This is necessary in view of the large number
 of theoretical papers\cite{Singh:arXiv0803.0429,Mazin:arXiv0803.2740,Ishibashi:arXiv0804.2963,Ma:arXiv0804.3370,Yildirim:arXiv0804.2252} 
 which appeared recently and often contain strong disagreements.
 This was partly studied by Mazin {\it et al.}.\cite{GGA-magnetism} 
Table \ref{As-mag-XC} summarizes the results: the magnetic moment on the Fe atom together with the total energy differences for each magnetic
 state studied here. Independent of the code or the XC functional used, the Q$_M$ AFM state is always found to be the ground state, which
 confirms our earlier report\cite{Yin:arXiv0804.3355}. The magnetic moment for both AFM orders are considerably larger than the ordered
moment reported from neutron diffraction and muon spin relaxation experiments, while the one for the FM order is much smaller.
 For this last case, FPLO7 gives zero which indicates no magnetism with both PZ and PW92 XC functional; 
Wien2K gives about 0.36 $\mu$$_B$ with GGA and PBE and 0.13 $\mu$$_B$ with PW92. 
It appears therefore that the magnetic moment of Fe for the same state with different XC functionals varies
by up to 0.5 $\mu$$_B$, which is unexpectedly large, although GGA is known to enhance magnetism.\cite{GGA-magnetism} 
The difference between FPLO7 and Wien2K in predicting the Fe magnetic moment for each state may explain  
 the total energy differences among them. 
Virtual doping (see subsection B) 
by 0.1 e$^-$/Fe enhances the Fe magnetic moment in the Q$_M$ AFM state but reduces it 
in the FM state for all the XC functionals used.

In the structural optimization (performed in the Q$_M$ state), 
FPLO7 with PW92 (LDA) functional gives reasonable c/a and z(La) in good agreement with experiment, 
but it predicted z(As) $\sim$ 0.139, which is 0.011 off the experimental value, about 0.1 \AA~in length.
However, Wien2K with PBE(GGA) XC functional gives an optimized z(As) $\sim$ 0.149, which agrees well with experimental z(As). Similar results
are found in the XFe$_2$As$_2$ family (X=Ba, Sr, Ca) too. It suggests that, GGA (PBE) XC functional optimizes the FeAs-based system much better
than LDA (PW92) XC functional. And GGA should have better performance in dealing with the structure (including c/a, equilibrium volume and z(As))
 under pressure of this FeAs family.
This is probably due to the layered structure of the FeAs family which results in large density gradient between layers,
thus GGA has better description of the potential.
 But in the meantime,
GGA (PBE) further overestimates the magnetic moment of Fe, which is already overestimated by LDA (PW92).

\begin{table}
\caption{Calculated magnetic moment of Fe, the amounts of total energy per Fe lie below nonmagnetic state of
FM, Q$_0$ AFM
and Q$_M$ AFM states from FPLO7 and Wien2K with different XC functionals of
LaOFeAs with experimental structure. Positive $\Delta$ EE means lower total energy than NM state. }
\label{As-mag-XC}
\begin{tabular}{|c|c|c|c|c|c|c|c|}
\hline
code & XC & \multicolumn{3}{|c|}{mag. mom. ($\mu$$_B$)} &
\multicolumn{3}{|c|}{$\Delta$ EE (meV/Fe)} \\
\hline
       &  & Q$_M$  & Q$_0$ & FM  & Q$_M$ & Q$_0$ & FM  \\
\hline
FPLO7  &PW92 & 1.87 & 1.72 & 0.00 & 87.2 & 24.6 & 0       \\
       &PZ   & 1.70 & 1.31 & 0.00 & 62.2 &  6.9 & 0      \\
\hline
WIEN2k &PW92  & 1.74 & 1.52 & 0.13 & 136.9 & 78.9 & 0     \\
       &GGA & 2.09 & 1.87 & 0.36 & 149.1 & 65.2 & 3.7    \\
       &PBE  & 2.12 & 1.91 & 0.37 & 158.1 & 70.2 & 4.5    \\
0.1 e$^-$ doped & PW92 & 1.86 & ---- & 0.08 & 125.2  & ---- & -0.5  \\
0.1 e$^-$ doped & GGA & 2.14 & ---- & 0.26 & 139.7  & ---- & -0.1  \\
0.1 e$^-$ doped & PBE & 2.16  & ---- & 0.27 & 149.6  & ---- & 2.1  \\
\hline
\end{tabular}
\end{table}

 \subsection{Effect of $z$(As) on the electronic structure of LaOFeAs}
Then we studied how the electronic structure of LaOFeAs depends on the value of z(As).
\begin{table}
\caption{Calculated magnetic moment of Fe, total energy relative to the nonmagnetic (ferromagnetic) states of NM/FM, Q$_0$ AFM
and Q$_M$ AFM of LaOFeAs with z(As)= 0.150 (experimental),0.145, and 0.139 (optimized)
from FPLO7 with PW92 XC functional.}
\label{As-mag}
\begin{tabular}{|c|c|c|c|c|c|c|c|}
\hline
z(As)  & \multicolumn{3}{|c|}{mag. mom. ($\mu$$_B$)} & \multicolumn{2}{|c|}{$\Delta$ EE (meV/Fe)}& \multicolumn{2}{|c|}{Fe 3d occ.$\#$} \\
\hline
        & Q$_M$     & Q$_0$     & FM   & FM-Q$_M$ & Q$_0$-Q$_M$ & maj. & min. \\
\hline
0.150   & 1.87     & 1.72     & 0.002  & 87.2        & 62.6  & 4.32 & 2.45  \\
0.145   & 1.70     & 1.41     & 0.000  & 60.5        & 54.0  & 4.24 & 2.55  \\
0.139   & 1.48     & 0.01     & 0.000  & 34.6        & 34.6  & 4.15 & 2.68  \\
\hline
\end{tabular}
\end{table}
Table \ref{As-mag} shows 
the difference between the experimental z(As)($\sim$ 0.150), the
optimized z(As) ($\sim$ 0.139) and a middle value of 0.145 when using FPLO7 with PW92 XC functional:
decreasing z(As) (reducing the Fe-As distance) rapidly reduces the differences in energy between the different
 magnetic orderings.
 At z(As) $= 0.145$, the magnetic moments of the Q$_M$ and Q$_0$ states are reduced significantly in comparison 
  with z(As) $= 0.150$, and the difference in energy has changed by around $20 \%$, indicating important changes
   in the electronic structure upon moving the As atom.
For z(As) $= 0.139$, the Q$_0$ AFM state has lost its moment (become the NM state),
 while the magnetic moment of the Q$_M$ state has decreased even more, with a changing rate 
 of $6.8$ $\mu$$_B$/\AA~, indicating strong magnetophonon coupling.\cite{Yin:arXiv0804.3355} 
Therefore, using the experimental or
  optimized value for the internal coordinate of As gives quite different results and might explain several
   of the discrepancies seen in the previously published works.
\begin{figure}[h]
\includegraphics*[angle=0,width=0.48\textwidth]{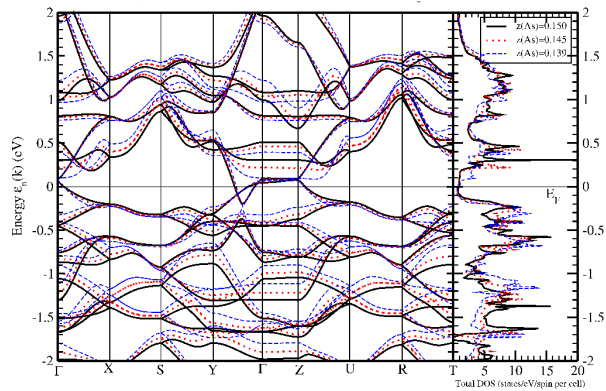}
\caption{
The bandstructure and total DOS of Q$_M$ LaOFeAs 
at ambient pressure computed for
z(As)=0.150, z(As)=0.145, z(As)=0.139.
\label{As-band}
}
\end{figure}
\begin{figure}[tbp]
\rotatebox{-90}
{\resizebox{6.0cm}{8.6cm}{\includegraphics{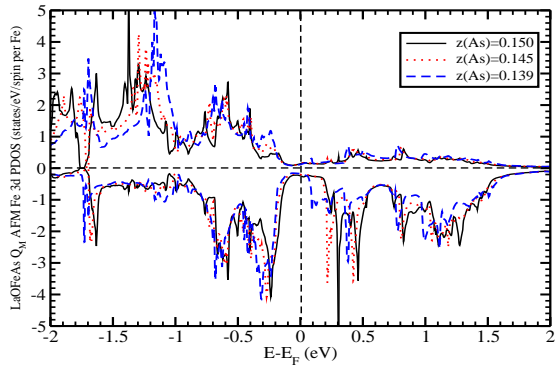}}}
\caption{ Plot of LaOFeAs Q$_M$ AFM Fe 3d PDOS at ambient pressure with
z(As)=0.150, z(As)=0.145, z(As)=0.139.}
\label{As-Fe-3d-PDOS}
\end{figure}
In Figures \ref{As-band} and \ref{As-Fe-3d-PDOS},
 we present the corresponding band structures, total densities
 of states, and partial densities of states calculated for different values of z(As).
Surprisingly, the band structure near E$_F$ referred to the common Fermi level barely changes when z(As) decreases.
Somewhat away from E$_F$, the bands below the Fermi level are pushed
 up in energy when z(As) is decreased, while the effect of the Fe-As distance on the bands above $\epsilon_F$ is less obvious,
  since they are pushed up or down depending on the direction of the Brillouin zone. For instance, along $\Gamma-X$ and $\Gamma-Z$
   they are pushed down, so that a decrease of the pseudogap is expected, as shown by Fig. \ref{As-band}.
The peaks of the DOS just above Fermi level move toward it when z(As) is reduced, while the DOS below the Fermi level is quite robust
with less changes. 
The important decrease of the magnetic moment of Fe when the Fe-As distance changes is understood by looking at the
 Fe-3d PDOS (Fig. \ref{As-Fe-3d-PDOS}) and the last column of table \ref{As-mag}. Although the number of Fe-3d electrons
  remains approximately constant, the number of spin up electron decreases, while the number of spin down electrons is
   increased when z(As) is reduced, which overall leads to a decrease of the magnetic moment.

\subsection{Effect of virtual crystal doping on the electronic structure of LaOFeAs}
Since superconductivity happens only in doped LaOFeAs, it is necessary to know how doping will
affect the underlying electronic structure and the character of each magnetic state. 
Using the experimental lattice parameters,
we performed virtual crystal doping calculations on LaOFeAs using Wien2K by changing the charge of O 
(doping with F) and La (doping with Ba, but simulating doping with Sr as well),
and the corresponding number of valence electrons.  The virtual crystal method is superior to
a rigid band treatment because the change in carrier density is calculated self-consistently
in the average potential of the alloy.

\begin{figure}[tbp]
\rotatebox{-90}
{\resizebox{6.0cm}{8.6cm}{\includegraphics{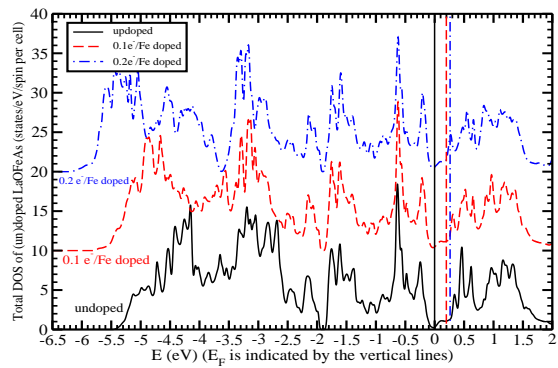}}}
\caption{ Plots of undoped, 0.1 and 0.2  electron-doped LaOFeAs Q$_M$ AFM total DOS (displaced upward 
consecutively by 10 units for clarity, obtained using the virtual crystal approximation. 
Referenced to that of the undoped compound,
the Fermi levels of 0.1 and 0.2 electron-doped DOS are shifted up by 0.20 eV and 0.26 eV, respectively.}
\label{O-doped-LaOFeAs-TDOS}
\end{figure}

There is only a weak dependence of the calculated Fe magnetic moment on the electron
doping level: 0.1 e$^-$/Fe doping enhances it from 2.12 ${\mu}$$_B$ to 2.16 ${\mu}$$_B$ (see Table \ref{As-mag-XC}). 
However, electron 
doping reduces the total energy difference (compared to NM) in both Q$_M$ AFM and FM states.
The main effect
of virtual crystal doping is to change the Fermi level position, in roughly a rigid band
fashion (see the caption of Fig. \ref{O-doped-LaOFeAs-TDOS} for more details). 
The band structures of 0.1, and 0.2 e$^-$/Fe doped LaOFeAs in the Q$_M$ AFM phase show only
small differences; the charge goes into states that are heavily Fe character and the small
change in the Fe $3d$ site energy with respect to that of As $4p$ states is minor.

Notably, the virtual crystal approximation continues to give strong magnetic states, whereas doping is observed
to degrade and finally kill magnetism and promote superconductivity. Thus the destruction of magnetism requires 
some large effect not considered here, such as strong dynamical spin fluctuations.

\subsection{Electric field gradients}
We have calculated the electric field gradients (EFG) of each atom in LaOFeAs, studying both the effects of doping
and of magnetic order.  The structure used for these calculations is
$a$=4.0355 \AA, $c$=8.7393 \AA, $z$(La)=0.142, $z$(As)=0.650, and the PBE(GGA) XC functional was used in the
Wien2K code. (PW92 (LDA) XC functional gives similar results and thus the results are not presented here.)
Since the EFG is a traceless symmetric 3$\times$3 matrix, only two of V$_{xx}$, V$_{yy}$, V$_{zz}$ 
are independent.  For cubic site symmetry, the EFG vanishes, hence the magnitude and sign of the EFG
reflects the amount and character of anisotropy of the charge density.
For the symmetries studied here, the off-diagonal components of the EFG tensor for
all the four atoms are zero.  For the
Q$_M$ AFM state, the V$_{yz}$ component calculated separately for each spin for La and As is not zero,
although the sum vanishes; the spin decomposition gives information about the anisotropy of the
spin density that is not available from measurements of the EFG.

As shown in Table
\ref{Fe-EFG} and Table \ref{As-EFG},
the EFGs of both Fe and As in NM and FM states are very similar
and they are doping
insensitive, except for Fe where the EFG is comparatively small (in tetrahedral symmetry, the EFG is identically zero).
Due to the breaking of the x-y symmetry in the Q$_M$ phase,
V$_{xx}$ is no longer equal to V$_{yy}$. 
In this case, the EFGs are quite different from 
those in the NM and
 FM states, which shows once more that the electronic structure in the Q$_M$ AFM
 order differs strongly from the ones of the NM and FM orders. Also, while hole doping (on the La
  site) and electron doping (on the O site) significantly change the EFG of Fe, the
   EFG of As is less affected.
Using nuclear quadrupolar resonance (NQR) measurement, Grafe {\it et al.}\cite{EFG-La} reported 
a quadrupole frequency $\nu$$_Q$=10.9 MHz and an asymmetry parameter $\eta$=0.1 of the As EFG in LaO$_{0.9}$F$_{0.1}$FeAs. 
This observation gives V$_{zz}$ $\sim$ 3.00 $\times$ 10$^{21}$ V/m$^2$, 
which agrees reasonably well with our result of 2.6 $\times$ 10$^{21}$ V/m$^2$ as shown in Table \ref{As-EFG} in the NM state.   
Upon $0.1$ electron or $0.1$ hole doping, the EFGs
   are modified in a similar way for As but differently for Fe.

\begin{table}
\caption{The EFG of Fe in LaOFeAs with NM, FM  and Q$_M$ AFM states at different doping levels 
from Wien2K with PBE(GGA) XC functional.
The unit is 10$^{21}$ V/m$^2$. }
\label{Fe-EFG}
\begin{tabular}{|c|c|c|c|c|c|c|c|}
\hline
Fe    &     & \multicolumn{3}{|c|}{V$_{xx}$} & \multicolumn{3}{|c|}{V$_{yy}$} \\
\hline
      & doping  & up    & dn     & total  & up    & dn     & total  \\
\hline
NM   & undoped  & 0.11  & 0.11    & 0.22   & 0.11   & 0.11   & 0.22  \\
     & 0.1h (La) & 0.21  & 0.21   & 0.42   & 0.21  & 0.21    & 0.42  \\
     & 0.1e (La) & 0.01  & 0.01   & 0.02   & 0.01  & 0.01    & 0.02  \\
     & 0.1e (O)  & 0.09  & 0.09    & 0.18   & 0.09  & 0.09    & 0.18  \\
\hline
FM   & undoped  & 0.51  & -0.30  & 0.21   & 0.51  & -0.30  & 0.21  \\
     & 0.1h (La) & 0.05  & 0.39  & 0.44   & 0.05  &  0.39  & 0.44   \\
     & 0.1e (La) & 0.31  & -0.21  & 0.10   & 0.31  & -0.21  & 0.10  \\
     & 0.1e (O)  & 0.31  & -0.20  & 0.11   & 0.31  & -0.20  & 0.11  \\
\hline
Q$_M$ & undoped & 0.22  & 0.03   & 0.25   & -1.11 & 0.54   & -0.57 \\
      &0.1h (La) & 0.60 & -1.13   & -0.43   & -1.15 & 1.04   & -0.11 \\
      &0.1e (La) & -0.55 & 1.00   & 0.45   & -1.05 & 0.24   & -0.81 \\
      & 0.1e (O) & -0.54 & 1.01   & 0.47   & -1.07 & 0.32   & -0.75 \\
      & 0.2e (O) & -0.82 & 1.17   & 0.35   & -1.02 & 0.52   & -0.50 \\
\hline
\end{tabular}
\end{table}

\begin{table}
\caption{The EFG of As in LaOFeAs with NM, FM  and Q$_M$ AFM states at different doping levels 
from Wien2K with PBE(GGA) XC functional.
The unit is 10$^{21}$ V/m$^2$. }
\label{As-EFG}
\begin{tabular}{|c|c|c|c|c|c|c|c|}
\hline
As    &     & \multicolumn{3}{|c|}{V$_{xx}$} & \multicolumn{3}{|c|}{V$_{yy}$} \\
\hline
      & doping  & up    & dn     & total  & up    & dn     & total  \\
\hline
NM   & undoped  & 0.69   & 0.69    & 1.38   & 0.69   & 0.69    & 1.38  \\
     & 0.1h (La) & 0.70  & 0.70   & 1.40   & 0.70  & 0.70    & 1.40   \\
     & 0.1e (La) & 0.65  & 0.65   & 1.31   & 0.65  & 0.65    & 1.31  \\
     & 0.1e (O)  & 0.66  & 0.66    & 1.32   & 0.66   & 0.66    & 1.32  \\
\hline
FM   & undoped  & 0.55  & 0.81   & 1.36   & 0.55  & 0.81   & 1.36  \\
     & 0.1h (La) & 0.58  & 0.68   & 1.26   & 0.58  & 0.68   & 1.26  \\
     & 0.1e (La) & 0.56  & 0.74   & 1.30   & 0.56  & 0.74   & 1.30  \\
     & 0.1e (O)  & 0.58  & 0.75   & 1.23   & 0.58  & 0.75   & 1.23  \\
\hline
Q$_M$ & undoped & -0.40 & -0.40  & -0.80  & 0.77  & 0.77   & 1.54 \\
      &0.1h (La) & -0.42 & -0.42  & -0.84  & 0.68  & 0.68   & 1.36 \\
      &0.1e (La) & -0.41 & -0.41  & -0.82  & 0.89  & 0.89   & 1.78 \\
      & 0.1e (O) & -0.40 & -0.40  & -0.80  & 0.91  & 0.91   & 1.82 \\
      & 0.2e (O) & -0.29 & -0.29  & -0.58  & 1.03  & 1.03   & 2.06 \\
\hline
\end{tabular}
\end{table}

\subsection{Effect of pressure on the electronic structure of LaOFeAs}

 Applying pressure is often used as a way to probe how the resulting effect on the electronic
structure impacts the superconducting critical temperature and other properties.
A strong pressure effect was shown experimentally for the members 
of the LaOFeAs family\cite{Takahashi2008,Lu-NJP-2008,Lorenz:arXiv0804.1582},
 since for example T$_c = 43$ K could be reached under pressure for LaO$_{1-x}$F$_x$FeAs, in case of optimal doping\cite{Takahashi2008}.
To begin to understand such observations, it is necessary to determine how the electronic structure 
of the parent compound LaOFeAs is changed by pressure.     

\begin{figure}[tbp]
\rotatebox{-90}
{\resizebox{6.0cm}{8.6cm}{\includegraphics{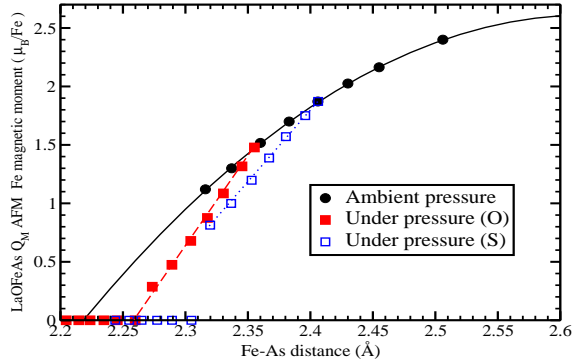}}}
\caption{ Plot of the magnetic moment of Fe atom in the Q$_M$ AFM state of LaOFeAs as a function of the Fe-As distance,
both at ambient pressure and under pressure.}
\label{As-mag-L}
\end{figure}

\begin{figure}[tbp]
\rotatebox{-90}
{\resizebox{6.0cm}{8.6cm}{\includegraphics{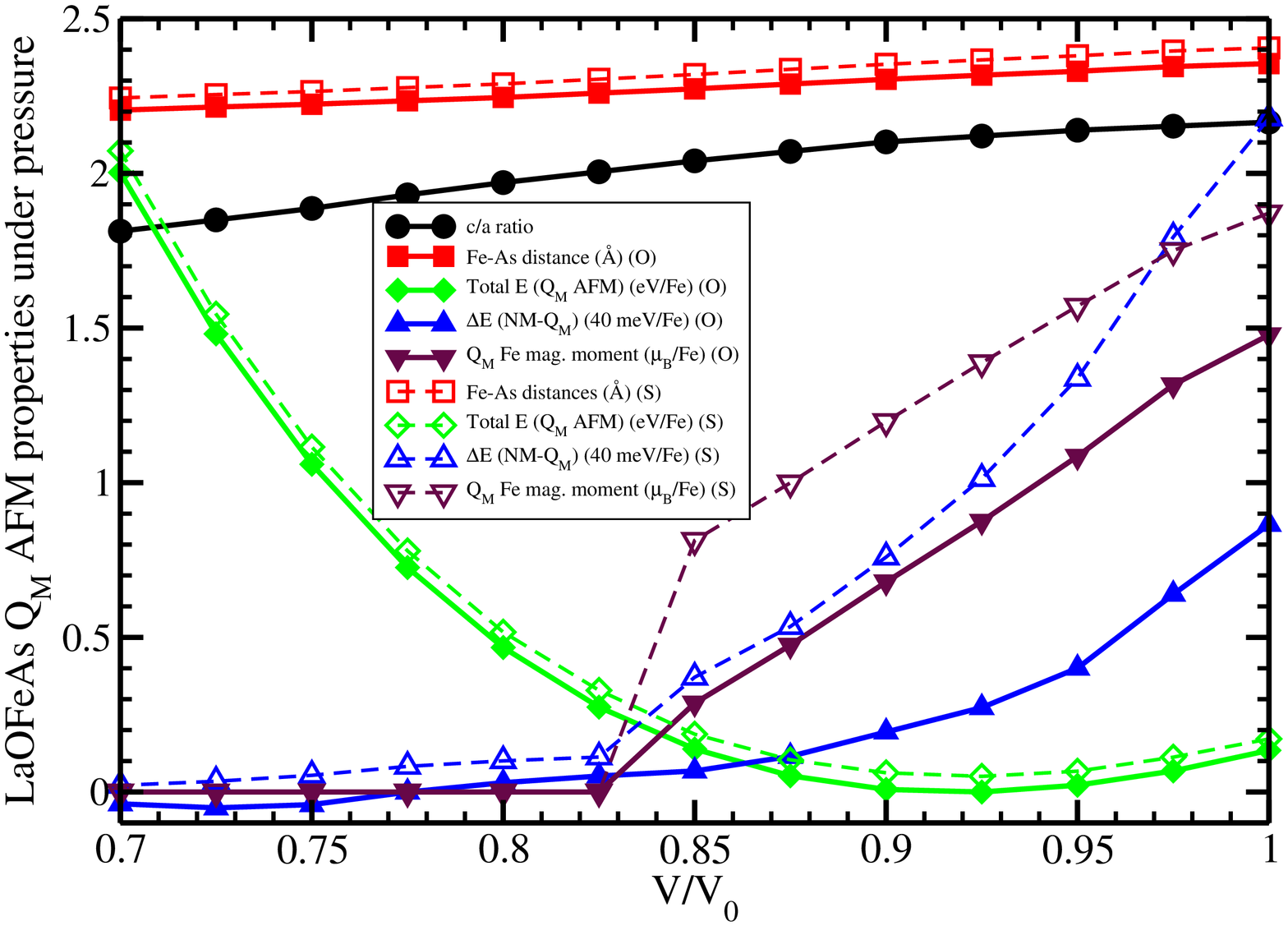}}}
\caption{ Plot of the optimized c/a ratio, the Fe-As distances (\AA),
the total energy of the Q$_M$ AFM state (eV), the total energy
differences between NM and Q$_M$ AFM state (EE(NM)-EE(Q$_M$ AFM) (40 meV/Fe),
the magnetic moment (${\mu}$$_B$) of the Q$_M$ AFM states as a function of V/V$_0$.}
\label{As-EE-V}
\end{figure}

In Fig. \ref{As-mag-L}, the magnetic moment of Fe in the Q$_M$ AFM phase versus Fe-As distance is presented.
Two different behaviours of the magnetic moment are observed. When $z$(As) is varied at constant volume
(zero pressure),the decrease of the magnetic moment of Fe
is parabolic.  When pressure is applied and all internal positions are optimized (hence $z$(As) changes)
the change is linear until the magnetic moment drops to zero.  This linear behavior is followed also
when the As height $z$(As) is shifted by 0.011 to compensate for the PW92 (LDA) error mentioned above.
Fig. \ref{As-EE-V} collects a number of results: the effect of pressure on the $c/a$ ratio, 
the Fe-As distance, the total energy,
 the difference in energy between NM and QM states, and the magnetic moment on Fe.
  Under pressure, the $c/a$ ratio, the Fe-As distance, and the magnetic moment of the Q$_M$ AFM state drop
linearly when volume is reduced. The PW92(LDA) predicts an equilibrium volume of 0.925 V$_0$; and the total energy
differences between NM and Q$_M$ AFM state gradually drops to zero at 0.78 V$_0$. 

\begin{figure}[tbp]
\rotatebox{-90}
{\resizebox{6.0cm}{8.6cm}{\includegraphics{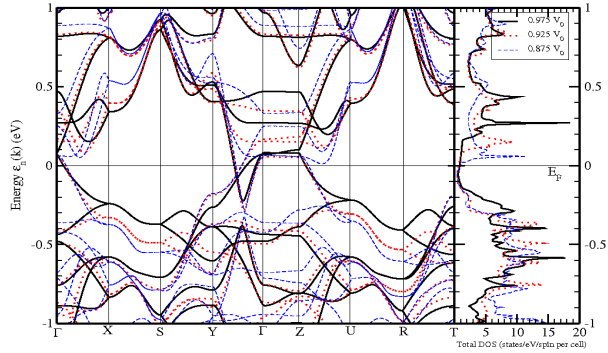}}}
\caption{The bandstructure and total DOS of Q$_M$ LaOFeAs computed for
0.975V$_0$, 0.925 V$_0$ and  0.875 V$_0$. z(As) has been shifted.}
\label{As-band-pressure}
\end{figure}

The effect of pressure on the band structure is shown in Fig. \ref{As-band-pressure}.
While the bands change positions under pressure, 
in the corresponding DOS (right panel of Fig. \ref{As-band-pressure}), the first peak above E$_F$ is moved towards the Fermi level when pressure
 is applied, but the DOS from -0.1 eV to E$_F$ is left almost unchanged by pressure. Therefore pressure should induces important changes
 in the superconducting properties of electron-doped LaOFeAs, while they should be less important for hole-doped LaOFeAs.

\begin{figure}[h]
\includegraphics*[angle=0,width=0.48\textwidth]{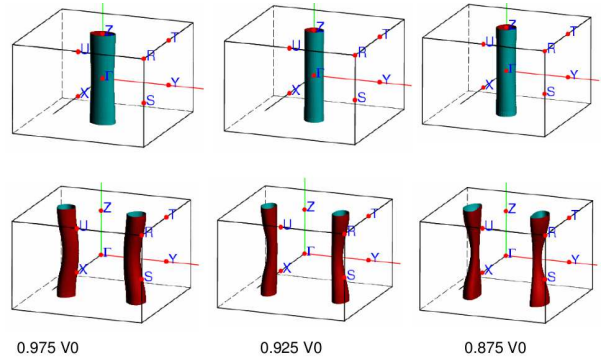}
\caption{
The Fermi surface of Q$_M$ LaOFeAs computed for
0.975V$_0$, 0.925 V$_0$ and  0.875 V$_0$. z(As) has been shifted.
\label{As-FS-pressure}
}
\end{figure}

The Fermi surface of Q$_M$ LaOFeAs computed for different values of the volume is presented in Fig. \ref{As-FS-pressure}.
The first sheet is an almost perfect cylinder along the $\Gamma-Z$ line, while the second sheet is made of
 two ellipsoidal cylinders with some k$_z$ bending. They appear to be very similar to the FS computed at ambient pressure\cite{Yin:arXiv0804.3355}.
 The pressure has almost no effect on the first sheet, but it enhances the distortion of the second sheet.

\section{\label{sect-pipipi} Investigation of the structural distortion and of the ($\pi,\pi,\pi$) magnetic order}

Magnetic $\vec q =$($\pi , \pi , \pi$) magnetic order in stoichiometric LaOFeAs (with 
magnetic cell being $\sqrt{2} \times \sqrt{2} \times 2$ the crystallographic cell)
was reported first  by de la Cruz et al.\cite{Cruz:arXiv0804.0795} using inelastic neutron scattering.  Ordering
takes place at $T_N = 135-140$ K and is preceded in temperature by a structural distortion occurring around $155$ K.
These transitions have since been confirmed by other groups.\cite{Nomura:arXiv0804.3569}
A similar structural distortion was found for NdOFeAs\cite{Fratini:arXiv0805.3992}, showing 
that the temperature of the structural phase
transition in this case is reduced by about 20 K in comparison with LaOFeAs. 
These phase transitions have been revisited\cite{McGuire:arXiv0806.3878}
using various experimental tools (heat capacity, ultrasound spectroscopy etc..).
 Although the magnetic ordering of FeAs layers along the $c$ axis is less likely to be
 crucial for the mechanism of superconductivity since the involved scale of energy
 is expected to be very weak is comparison with the intralayer ordering, its study is necessary
 to understand the complete system. For the same reason, and even if there are strong 
 indications that it does not happen
 in the case of F-doped LaOFeAs, it is interesting to see whether the structural distortion 
 of pure LaOFeAs can be reproduced
 by ab-initio calculations, and what the corresponding electronic structure looks like. The results 
 we present in this section were obtained using the VASP code with the PBE(GGA) functional\cite{PBE}.

\subsection{($\pi,\pi,0$) structural order}
\begin{figure}[h!]
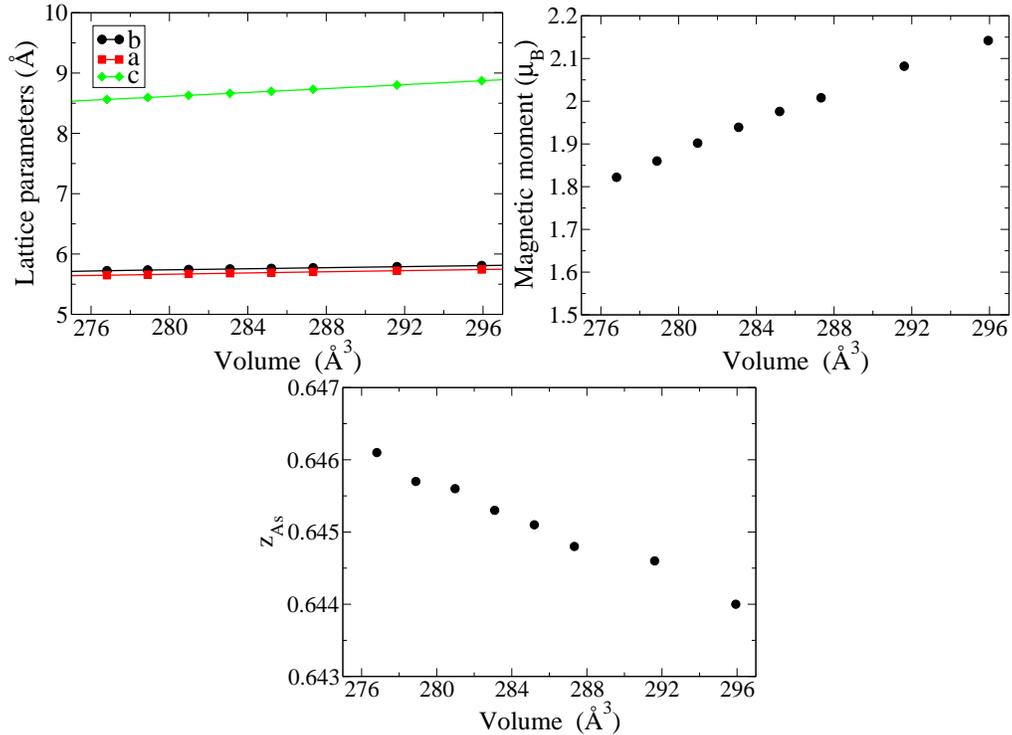

\includegraphics*[angle=0,width=0.40\textwidth]{abc.eps}
\includegraphics*[angle=0,width=0.40\textwidth]{mag.eps}
\includegraphics*[angle=0,width=0.40\textwidth]{zas.eps}
\caption{
Upper plot: Lattice parameters versus volume.
Middle plot: magnetic moment versus volume.
Lower plot: z$_{As}$ versus volume.
\label{fig:full-relax}
}
\end{figure}

The structural transformation\cite{Cruz:arXiv0804.0795,Nomura:arXiv0804.3569} 
changes the $\sqrt{2} \times \sqrt{2}$ cell (with four iron atoms; full lines in Fig. \ref{fig:afm})
  from tetragonal (space group $P4/nmm$) to orthorhombic (space group $Cmma$) 
  or equivalently for the primitive cell (with two iron atoms; dashed lines in Fig. \ref{fig:afm}) from tetragonal (space group $P4/nmm$) to 
   monoclinic (space group $P112/n$).   
To simplify our study, the cell doubling along the $c$ axis due to magnetic ordering is neglected for this study, i.e. we consider 
   only the ($\pi \pi 0$) order. 
 We have performed a relaxation (shape of the cell as well as atom positions) of LaOFeAs for different volumes, the results
    being presented in Fig. \ref{fig:full-relax}.  The calculated equilibrium lattice parameters as
     well as the internal atomic positions are reported in Table \ref{tab:eq}, together with available experimental data.
     The overall agreement is satisfactory, the length of the $a$ and $b$ lattice cell vectors being slightly overestimated
      by our calculations, while the value of $c$ is slightly underestimated.
      The value of $|\delta |$ (the monoclinic distortion angle) is overestimated by our calculations,
     but the very small distortion and very small energy difference makes this difference understandable.
  The important point is that
        ab-initio calculations are indeed able to reproduce the structural instability of LaOFeAs.

      As for the atom positions within the cell, the agreement is good for the positions of La, O, and Fe but
        is less satisfying for the internal position $z$(As) of arsenic.  The difficulty
       concerning the position of As has been reported by us previously\cite{Yin:arXiv0804.3355} and is related to the
        strong magnetophonon coupling that occurs in this compound.
      In Fig. \ref{fig:full-relax}, we present the corresponding lattice parameters (upper plot);
      magnetic moment (middle plot); and internal coordinate of As ( z$_{As}$) (lower plot), versus volume for LaOFeAs. 
      The range of pressure covered goes roughly from $-2.5$ GPa to $2.5$ GPa.
      By fitting the E-V data (not shown here) to a Birch-Murnaghan equation of state (EOS), we find LaOFeAs to have a 
      bulk modulus of B$_{0} = 75$ GPa
        and a bulk modulus derivative B$^{'}_{0} = 4.1 $. Also, from the upper plot of Fig. \ref{fig:full-relax},
       we predict that LaOFeAs is more
     compressible along the $c$ axis than along the $a$ and $b$ axes, a common characteristic of layered materials.

 More important is the dependence of the magnetic moment on the volume (middle plot of Fig. \ref{fig:full-relax}).
     This dependence has two origins: the first one is the usual dependence of the magnetic moments on the volume change, but
      in LaOFeAs, the magnetic moment on Fe is known\cite{Yin:arXiv0804.3355} to be strongly dependent on the internal coordinate of 
      As which changes with applied pressure (lower plot of Fig. \ref{fig:full-relax}).

\begin{table}
\caption{Left and middle columns: the structure parameters of LaOFeAs in its low-temperature phase as obtained from
x-ray\cite{Nomura:arXiv0804.3569} and neutron\cite{Cruz:arXiv0804.0795} studies, as reported
by Yildirim\cite{Yildirim:arXiv0805.2888}. Right column: results from calculations obtained after a full relaxation
 of a $\sqrt{2} \times \sqrt{2}$ cell with a ($\pi \pi 0$) magnetic order.
  $a$, $b$, and $c$ are the lattice parameters, $|\delta |$ is the monoclinic distortion angle of
   the primitive cell, and La(z), As(z), O(z), and Fe(z) are the internal coordinate of the corresponding atom.
\label{tab:eq}
}
\begin{center}
\begin{tabular}{|c||c|c||c|c|} \hline 
 &\multicolumn{2}{|c||}{X-ray (120 K) } & Neutron (4K) & Calcs. \\
 & $\sqrt{2} \times \sqrt{2}$ &  Primitive  &  Primitive   & $\sqrt{2} \times \sqrt{2}$ \\ \hline\hline
$a$ & 5.68262 \AA & 4.02806 \AA & 4.0275 \AA& 5.69 \AA\\
$b$ & 5.71043 \AA & 4.02806 \AA & 4.0275 \AA& 5.76 \AA\\
$c$ & 8.71964 \AA & 8.71964 \AA & 8.7262 \AA& 8.70 \AA\\
$|\delta |$ &  \multicolumn{2}{|c||}{  $ 0.2797^{\rm o}$} & $0.279^{\rm o}$  & $0.69^{\rm o}$\\
La(z) & \multicolumn{2}{|c||}{0.14171} &  0.1426  & 0.1418 \\
As(z) & \multicolumn{2}{|c||}{0.65129} & 0.6499 & 0.6451 \\
O(z)  & \multicolumn{2}{|c||}{0} & -0.0057 & 0.0  \\
Fe(z) & \multicolumn{2}{|c||}{0.5}  &  0.5006  & 0.5  \\ \hline\hline
\end{tabular}
\end{center}
\end{table}

The structural distortion has been addressed by Yildirim,\cite{Yildirim:arXiv0804.2252,Yildirim:arXiv0805.2888}
who approached the question differently and obtained different results.  While our value of the Fe moment 
is close to that for the
undistorted structure as would be expected, the moment reported by Yildirim is $0. 48~\mu_B$ per Fe atom.
We checked carefully the possible existence of such a magnetic solution,
 but our calculations appears to be robust, with the magnetic moment of Fe being around $2~\mu_B$.
As a result of the different magnetic moment, his computed DOS (see Fig. $5$ in Ref. \onlinecite{Yildirim:arXiv0804.2252}) 
also is different.
  Together with an experimental study, Nomura et al.\cite{Nomura:arXiv0804.3569}
  reported ab-initio calculations on LaOFeAs for both the tetragonal and orthorhombic structures, and found almost vanishing magnetic moments,
   which correspond to a non-magnetic ground-state. In our case, such a state is higher in energy by about $140$ meV per Fe atom for the fully relaxed
    structure, and therefore can safely be ruled out as being the true ground-state of LaOFeAs.
The differences in calculated values that we have noted reflect an unusual sensitivity to details (structure, method, XC functional).

 \begin{figure}[h]
\includegraphics*[angle=0,width=0.48\textwidth]{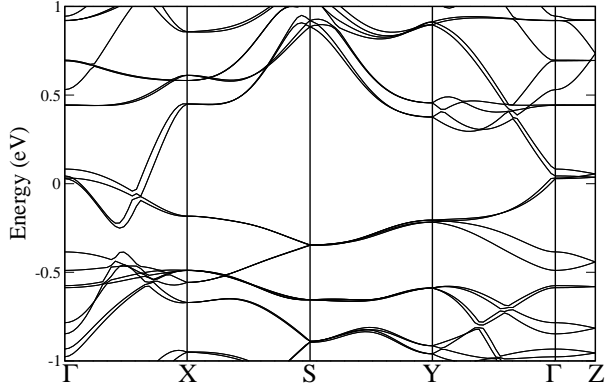}
\caption{
 The bandstructure of LaOFeAs along high-symmetry directions in the case of a ($\pi \pi \pi$)
  magnetic order for the distorted (orthorhombic) $\sqrt{2} \times \sqrt{2} \times 2$ cell.
 The high symmetry points are defined as $\Gamma$ $(0,0,0)$; X $(0.5,0,0)$; S $(0.5,0.5,0)$; Y $(0,0.5,0)$; and Z $(0,0,0.5)$, in
  terms of reciprocical lattice vectors.
\label{bands-pipipi}
}
\end{figure}

\subsection{($\pi,\pi,\pi$) magnetic order}
We turn now to the investigation of LaOFeAs taking into account both the true ($\pi,\pi,\pi$) magnetic order and
  the structural distortion.
  In this case, we have used the experimental structural data provided by de la Cruz et al.
  \cite{Cruz:arXiv0804.0795}. As in the case of the ($\pi,\pi,0$) order, there are two possible magnetically ordered states.
    Only one gives the ($\pi,\pi,\pi$) order to be the
    ground state versus the ($\pi,\pi,0$) order, and by only few meV per Fe atom. This small energy difference is near
the limit of precision of our calculations, but appears to confirm
     the sign of the very weak magnetic interaction along the $c$ layers.

  The corresponding band structure is shown in Fig. \ref{bands-pipipi}. Due to the doubling of the cell along the $c$ axis,
   there are now four bands crossing the Fermi level (see Fig. $3$
    of Ref. \onlinecite{Yin:arXiv0804.3355}).
  Along S-Y there are tiny
  splittings around -0.25 and -0.6 eV as well as along $\Gamma$-X and $\Gamma$-Y, indicating the magnitude of interlayer coupling.
  In particular, the splitting is particularly large for one pair of bands just above E$_F$ at $\Gamma$. Along X-S-Y, 
  the bands nearest the Fermi level are hardly split at all.
  Overall, the band structure retains the essential features noticed before\cite{Yin:arXiv0804.3355},
  namely a pseudogap separating bonding and antibonding states over much of the zone, 
  together with dispersive bands crossing the Fermi level
   along only one of the two in-plane directions ($\Gamma-X$, with our choice of axes).

 The total and partial densities of states are very similar to the ones in the case of a ($\pi \pi 0$) magnetic order and won't be shown here;
  but we notice that the rough electron/hole symmetry in view of the study of doped (superconducting) materials is preserved.
  Also, our calculated Fermi surface (not shown here), made of four sheets, is very similar to the one presented
    previously\cite{Yin:arXiv0804.3355} for the ($\pi,\pi,0$) order and folded back along k$_z$: it has two sheets 
  along the $\Gamma-Z$ direction which are almost perfectly cylindrical, while the two other sheets are more distorted, 
  but still showing a strong two-dimensional character.

\section{Role of the Pnictogen Atom}
As mentioned at the beginning of Section I of this paper, 
LaOFeAs and LaOFeP are isostructural and isovalent, but they have quite different properties: 
LaOFeAs is Q$_M$ AFM ordered below T$_N$=150 K and not superconducting, 
while LaOFeP is a T$_c$=2.5 K superconductor\cite{Kamihara2006} without magnetic order. 
Also, they have completely different response to doping:
either electron or hole doping will destroy the Q$_M$ AFM ordering in LaOFeAs and make it 
superconducting with T$_c$ over 26 K\cite{Kamihara2008} (43 K under pressure\cite{Takahashi2008}),
 while in LaOFeP, doping changes the critical temperature less significantly to only 9 K\cite{Kamihara2006}.
A deeper understanding of the differences of the electronic structure of these two compounds can provide insight
into the competition between magnetic ordering and superconductivity.  For similar reasons,
 the related compounds LaOFeN and LaOFeSb (although not studied experimentally yet) are potentially of high
  interest, so we also provide predictions for their electronic structure.

Table \ref{Pn-lattice} displays the experimental structure parameters for LaOFeP\cite{Kamihara2006} and LaOFeAs\cite{Kamihara2008}
 as well as the predicted structure for LaOFeN and LaOFeSb
after optimization (see below for calculation details).
As a result of the increasing size of the pnictogen atom, the Fe-Pn length changes. 
 In particular, the Fe-Pn distance is consistent with
the sum of the covalent radii of Fe and Pn, which reflects the covalent bonding nature
between Fe and Pn atoms in this family. The slight increase of the La-O distance through the series
 is just a size effect related to the expansion of the volume .

\begin{table}[th]
\begin{centering}
\begin{tabular}{|c|c|c|c|c|c|c|c|c|}
\hline
Pn & a (\AA) & c (\AA)  & c/a     & z(La) & z(Pn) & La-O         & Fe-Pn         & Sum       \\
\hline
N  & 3.6951    & 8.0802     & 2.187  & 0.170 & 0.109 & 2.302       & 2.047        & 2.00       \\
P  & 3.9636    & 8.5122     & 2.148  & 0.149 & 0.134 & 2.352       & 2.286        & 2.31       \\
As & 4.0355    & 8.7393     & 2.166  & 0.142 & 0.151 & 2.369       & 2.411        & 2.44       \\
Sb & 4.1626    & 9.3471     & 2.246  & 0.127 & 0.171 & 2.396       & 2.624        & 2.62       \\
\hline
\end{tabular}
\end{centering}
\caption{Structural parameters of LaOFePn (Pn = N, P, As, or Sb), as obtained
 experimentally for LaOFeP\cite{Kamihara2006} and LaOFeAs\cite{Kamihara2008} or
  from our calculations for LaOFeN and LaOFeSb.
Length units are in~\AA, $z$(La) and $z$(Pn) are the internal coordinate of the lanthanum atom
 and the pnictide atom, and
``Sum" means the sum of Fe covalent radius and the Pn covalent radius, which is quite close to the
calculated value in all cases.}
\label{Pn-lattice}
\end{table}

\begin{table}
\begin{tabular}{|c|c|c|c|c|c|}
\hline
Pn  & \multicolumn{3}{|c|}{mag. mom. ($\mu$$_B$)} & \multicolumn{2}{|c|}{$\Delta$ EE (meV/Fe)} \\
\hline
        & Q$_M$     & Q$_0$     & FM      &    FM-Q$_M$ & Q$_0$-Q$_M$ \\
\hline
N       & 1.63     & 0.80     & 0.027  & 41.0        & 40.0         \\
P       & 0.56     & ----     & 0.087  & 1.6         & ----        \\
As      & 1.87     & 1.72     & 0.002  & 87.2        & 62.6        \\
Sb      & 2.47     & 2.43     & 0.000  & 293.8       & 82.4        \\
\hline
\end{tabular}
\caption{Calculated magnetic moment of Fe, total energy relative to the nonmagnetic (ferromagnetic) states of Q$_0$ AFM,
and Q$_M$ AFM states of LaOFePn from FPLO7 with PW92 XC functional.
}
\label{Pn-mag}
\end{table}

The values of the Fe magnetic moment for LaOFePn with FM/NM, Q$_0$ AFM, Q$_M$ AFM states,
and their total energy differences are presented in Table \ref{Pn-mag}.
Apart from LaOFeP, all the members of the LaOFePn family studied here
 have a large Fe magnetic moment in the Q$_M$ AFM state,
 the corresponding total energy being significantly lower than the ones corresponding to FM/NM state.  

\subsection{LaOFeP}

LaOFeP was the first member of the iron-oxypnictide family to be reported to be superconducting\cite{Kamihara2006}.
The corresponding electronic structure was studied by Leb\`egue using ab-initio calculations\cite{Lebegue2007}, but
 considering only a non-magnetic ground-state. Since then LaOFeP has been studied using various experimental tools:
  by using photoemission\cite{Ishida:arXiv0805.2647,Malaeb:arXiv0806.3860,Lu:arXiv0807.2009}, it was shown that 
  the Fe 3$d$ electrons are itinerant, and that there is no pseudogap in LaOFeP. Also, magnetic measurements revealed\cite{kamihara:214515,McQueen:arXiv0805.2149}
   that LaOFeP is a paramagnet, while electron-loss spectroscopy\cite{che:184518} implied a significant La-P hybridization.
    The absence of long-range order in LaOFeP was confirmed by M\"ossbauer spectroscopy
\cite{Tegel:arXiv0805.1208} and it was proposed that LaOFeP and doped LaOFeAs could have different mechanisms to drive the superconductivity
 in these compounds. Also, further theoretical studies were performed\cite{kamihara:214515,che:184518,Lu:arXiv0807.2009} but without
 studying all the possible magnetic states.

In our calculations, we find that for FM order Fe has a weak
magnetic moment of about 0.09 $\mu$$_B$, with a total energy very close to the NM one; this result is much like 
what is found in LaOFeAs.
A remarkable difference is that the Q$_0$ AFM state cannot be obtained. 
 However, we found the Q$_M$ AFM state to be the lowest in energy, 
 but only by about 1.6 meV/Fe, which is about two orders of magnitude less than in LaOFeAs.  LaOFeP, therefore, 
presents the situation where all
of the three possible magnetic states are all very close in energy to the nonmagnetic state,
   in contrast with LaOFeAs for which the Q$_M$ AFM order was clearly the ground state. Thus LaOFeP is surely
   near magnetic quantum criticality.

\begin{figure}[tbp]
\rotatebox{-90}
{\resizebox{5.0cm}{8.6cm}{\includegraphics{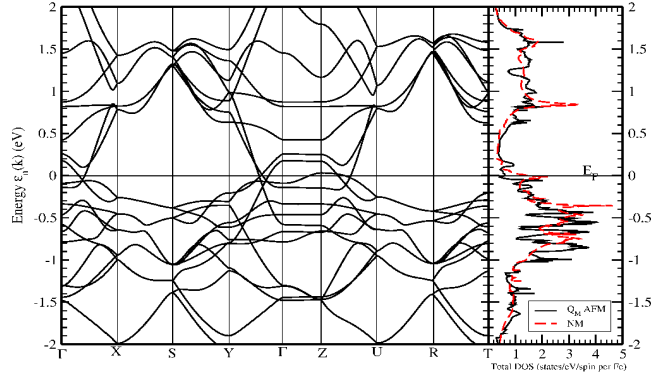}}}
\caption{Plot of LaOFeP band structure in Q$_M$ AFM state and total DOS in both Q$_M$ AFM and NM statesg
 at ambient conditions with experimental lattice parameters.}
\label{P-band-DOS-QM}
\end{figure}

The band structure of Q$_M$ AFM LaOFeP is displayed in Fig. \ref{P-band-DOS-QM}
together with total DOS for both Q$_M$ AFM and NM states. The band structure of Q$_M$ AFM LaOFeP is quite different from that
of LaOFeAs with the same Q$_M$ order, with the most significant differences along $\Gamma$-X, $\Gamma$-Y and $\Gamma$-Z lines. 
The difference is because the breaking of the $x-y$ symmetry
is much smaller in the Q$_M$ AFM LaOFeP compared to LaOFeAs, because the calculated Fe moment is only
0.56 $\mu$$_B$ in LaOFeP (it is 1.87 $\mu$$_B$ in LaOFeAs with the same calculational method).
The corresponding DOS is also different from that of LaOFeAs:
 there is structure within the  pseudogap around Fermi level in LaOFeP (See Fig. \ref{P-band-DOS-QM}). 
The difference in total DOS at E$_F$ is significant: it is only 0.2 states/eV/spin per Fe for LaOFeAs, 
but it is 0.6 states/eV/spin per Fe for LaOFeP. In the NM state of LaOFeP, it is even larger with
1.6 states/eV/spin per Fe. The DOS of Q$_M$ AFM LaOFeP is fairly flat from the Fermi level (set to 0.0 eV) 
to 0.6 eV, so that electron doping of 
LaOFeP will increase the Fermi level, but will hardly change N(E$_F$) (in a rigid band picture).

An important consequence is that there will be no expected enhancement of T$_C$ coming from N(E$_F$) upon electron doping.
 In order to see a significant increase of N(E$_F$) in Q$_M$ AFM LaOFeP,
an electron doping level of at least 1.2 e$^-$/Fe is required, which seems unrealistically large based on the
current experimental information.
This conclusion remains valid in the case of NM LaOFeP, since apart from a peak around Fermi level,
 the DOS is about the same as for the Q$_M$ AFM state.
Again, the behavior is quite different from the one of Q$_M$ AFM LaOFeAs: 0.1 e$^-$/Fe doping will increase its N(E$_F$)
by a factor of 6: from 0.2 states/eV/spin per Fe to 1.2 states/eV/spin per Fe. 

\begin{figure}[ht]
\subfigure[FS1]{
{\resizebox{3.6cm}{3.6cm}{\includegraphics{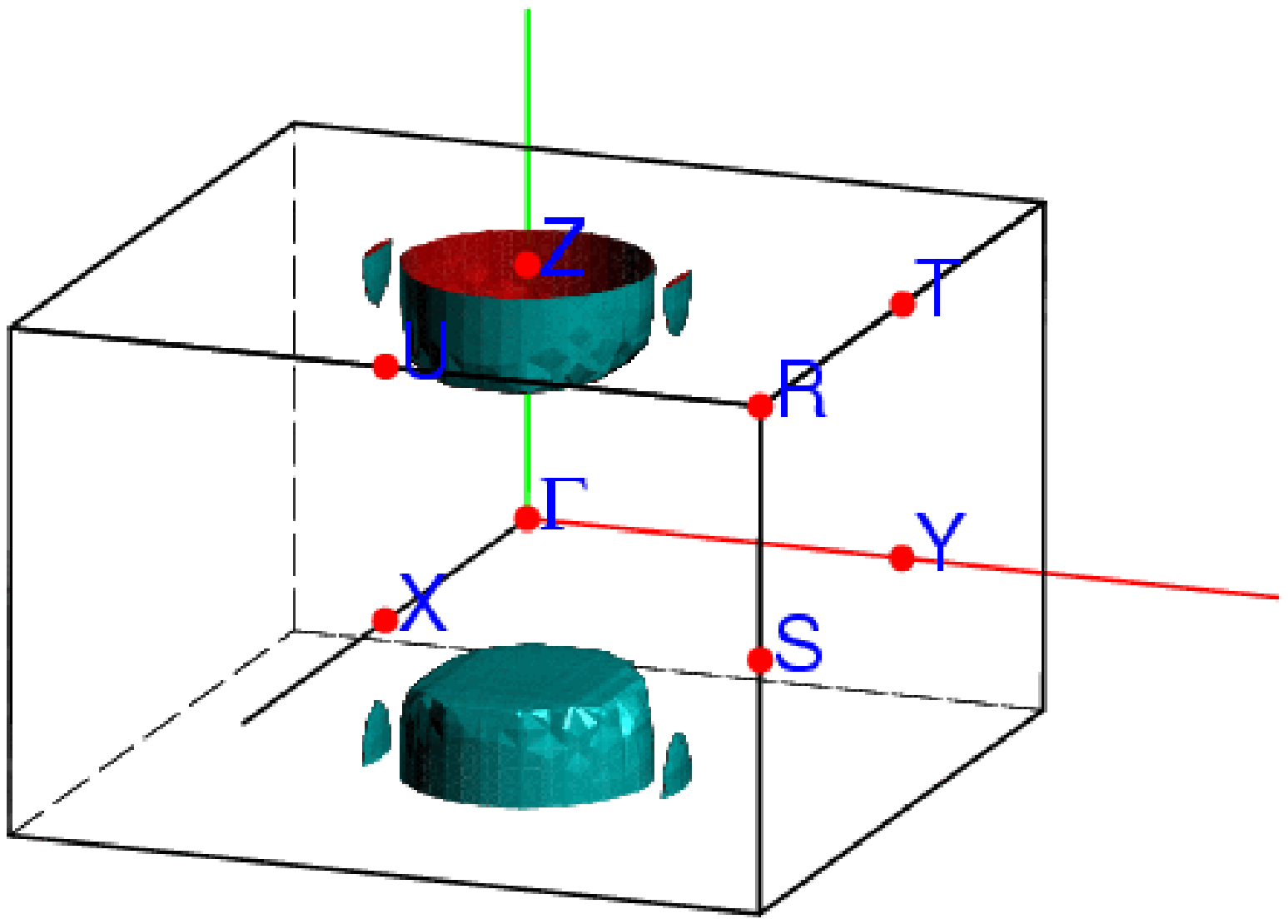}}}
\label{fig:P-QM-FS1}
}
\subfigure[FS2]{
{\resizebox{3.6cm}{3.6cm}{\includegraphics{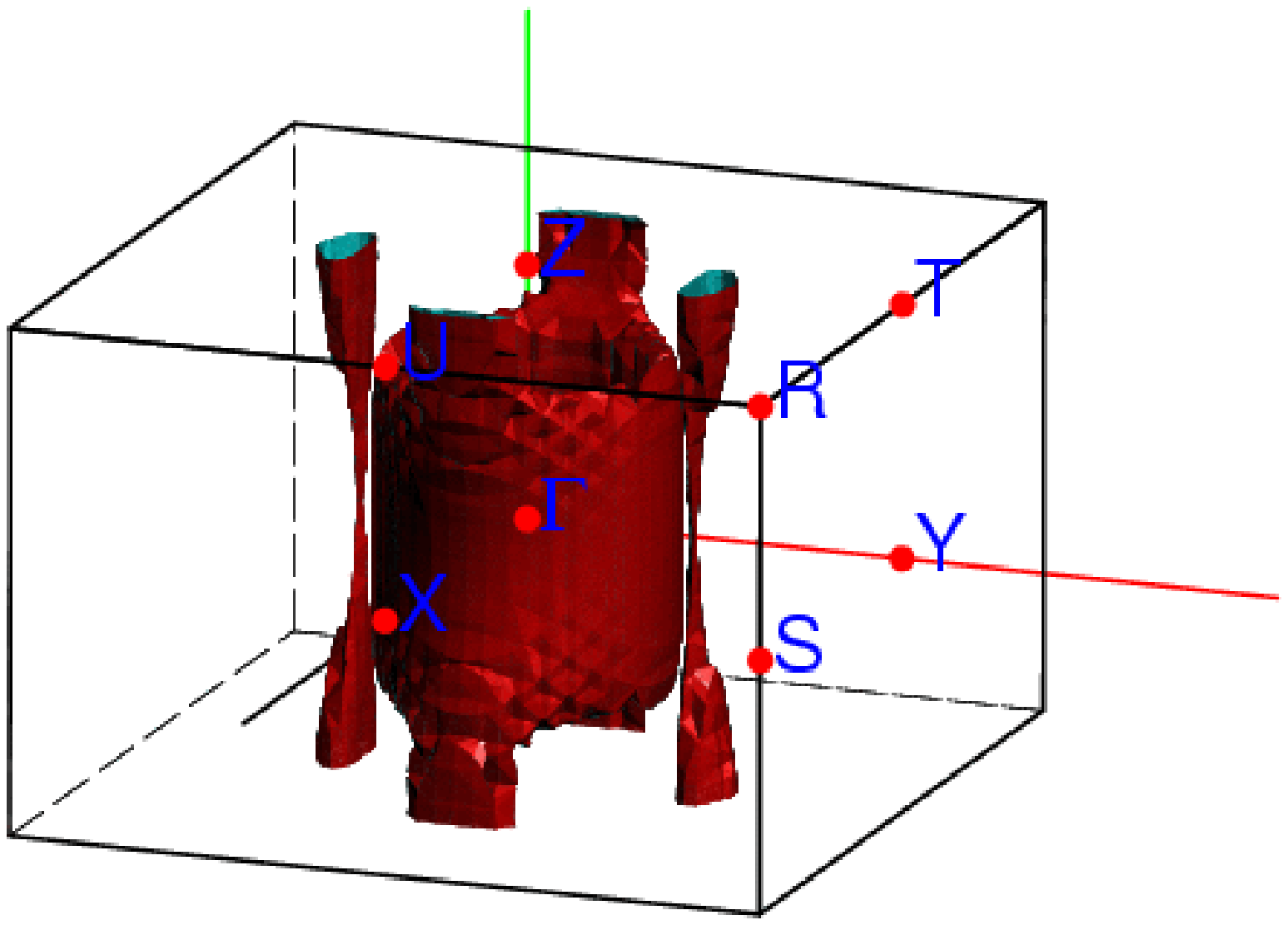}}}
\label{fig:P-QM-FS2}
}
\caption[ ]
{Fermi surface of Q$_M$ AFM LaOFeP, showing the very strong differences compared to LaOFeAs.
\label{fig:P-QM-FS}
}
\end{figure}

The Fermi surface of Q$_M$ AFM LaOFeP is shown in Fig. \ref{fig:P-QM-FS}. Compared to the Fermi surface of Q$_M$ AFM 
LaOFeAs presented earlier by Yin {\it et al.}\cite{Yin:arXiv0804.3355},
the piece enclosing the $\Gamma$-Z line (containing holes) increases in size and its $x-y$ cross section 
becomes more circular rather than elliptic. There is another piece (absent in LaOFeAs) also
enclosing the $\Gamma$-Z line with the same shape but larger in size and containing electrons instead of holes. 
The two symmetric electron-type pieces of Fermi surface lying along $\Gamma$-Y direction 
in LaOFeAs reduces a lot in size in LaOFeP but it has two additional similar pieces
lying along $\Gamma$-X direction. In LaOFeP, it has one more hole-type piece of Fermi surface surround Z point, which is a small cylinder.
It is, understandably, quite different from the Fermi surface of NM LaOFeP presented earlier\cite{Lebegue2007}.

Therefore, while they are isostructural and significantly covalent, LaOFeP and LaOFeAs present quite 
important differences in their respective electronic 
 structures. These differences must form the underpinning of any explanation of 
why LaOFeP is superconducting with a T$_c$ which is almost electron-doping independent, while pure
  LaOFeAs is not superconducting and becomes so only upon doping.

\subsection{LaOFeSb}

Since the experimental crystal structure of LaOFeSb is not reported yet, we conducted calculations
to obtain the structure. The procedure we used is the following:
starting from the experimental
volume V$_0$ of LaOFeAs (but with As replaced by Sb), we first
optimized $c/a$, $z$(La) and $z$(Sb). Then we chose a higher volume and again optimized the parameters, 
finally finding the volume that has the lowest total energy. 
Using this scheme, the optimized volume is
1.046 V$_0$ while for LaOFeAs the equilibrium volume is about 0.919 V$_0$. Assuming that PW92 overbinds equally 
 for LaOFeSb as for LaOFeAs, the experimental equilibrium volume for LaOFeSb should be 1.046/0.919=1.138 V$_0$.
Therefore, we performed calculations for a range of volume from V $=$ V$_0$ to  V $= 1.150$ V$_0$, the
 corresponding structural parameters being presented in Table \ref{Sb-lattice-V}.
\begin{table}
\begin{tabular}{|c|c|c|c|c|c|}
\hline
V/V$_0$        & a ($\AA$) & c ($\AA$) &  $c/a$   & z(La) & z(Sb) \\
\hline
1.000          & 4.092  & 8.500  & 2.077 & 0.137 & 0.165 \\
1.050          & 4.118  & 8.812  & 2.140 & 0.133 & 0.163 \\
1.100          & 4.141  & 9.131  & 2.205 & 0.129 & 0.161 \\
1.125          & 4.155  & 9.274  & 2.232 & 0.128 & 0.160 \\
1.138          & 4.163  & 9.347  & 2.245 & 0.127 & 0.160 \\
1.150          & 4.169  & 9.418  & 2.259 & 0.126 & 0.159 \\
\hline
\end{tabular}
\caption{Optimized structure parameters for LaOFeSb at several volumes. The accuracy for $c/a$ is within 0.3$\%$,
and within 0.8$\%$ for $z$(La) and $z$(Sb).}
\label{Sb-lattice-V}
\end{table}

\begin{figure}[tbp]
\rotatebox{-90}
{\resizebox{5.0cm}{8.6cm}{\includegraphics{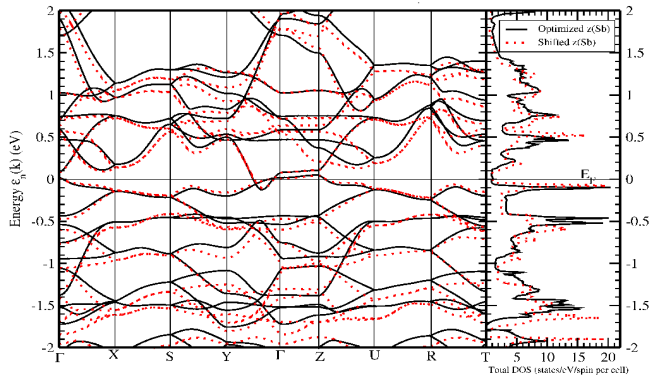}}}
\caption{Plot of Q$_M$ AFM LaOFeSb band structure and total DOS at 1.138 V$_0$ with both optimized and shifted z(Sb).}
\label{Sb-band-DOS-QM}
\end{figure}

Since for LaOFeAs in the Q$_M$ AFM phase PW92 underestimated $z$(As) by 0.011 at its experimental volume, we
 corrected $z$(Sb) by adding 0.011 to the optimized $z$(Sb) (we refer to this position at the ``"shifted $z$(Sb)").
Both for the NM and Q$_M$ AFM case, there are very small differences near E$_F$ between the optimized $z$(As) and shifted $z$(As)
in the band structure and DOS, as seen in Fig \ref{Sb-band-DOS-QM}.
However, shifting $z$(Sb) induces important changes in the energy differences between NM and Q$_M$ AFM
 states, as shown in Table \ref{Sb-mag-th}.
Also, the magnetic moment of Fe, and the energy differences
among NM/FM, Q$_0$ AFM and Q$_M$ AFM are strongly dependent on the volume.
With decreasing volume, 
 the difference in energy between the different magnetic states decreases quickly.

\begin{table}
\begin{tabular}{|c|c|c|c|c|c|}
\hline
V/V$_0$         & \multicolumn{3}{|c|}{mag. mom. ($\mu$$_B$)} & \multicolumn{2}{|c|}{$\Delta$ EE (meV/Fe)} \\
\hline
               & Q$_M$     & Q$_0$     & FM     & FM-Q$_M$    & Q$_0$-Q$_M$ \\
\hline
1.000          & 1.58      & 1.12      & 0.36   & 60.1        & 60.1     \\
1.050          & 1.87      & 1.74      & 0.44   & 95.6        & 68.0     \\
1.100          & 2.09      & 2.00      & 0.00   & 147.6       & 70.7    \\
1.125          & 2.17      & 2.10      & 0.00   & 172.6       & 71.8   \\
1.138          & 2.23      & 2.16      & 0.00   & 190.6       & 72.5   \\
1.150          & 2.26      & 2.19      & 0.00   & 199.0       & 72.0   \\
\hline
1.050          & 2.17  & 2.08  & 0.72  & 158.1 & 78.0 \\
1.100          & 2,35  & 2.00  & 0.00  & 223.8 & 80.8 \\    
1.125          & 2.42  & 2.37  & 0.00  & 271.6 & 81.8       \\
1.138          & 2.47  & 2.42  & 0.00  & 293.8 & 82.4      \\
1.150          & 2.49  & 2.45  & 0.00  & 287.6 & 82.1      \\
\hline
\end{tabular}
\caption{Calculated magnetic moment of Fe, total energy relative to the nonmagnetic (ferromagnetic) states of Q$_0$ AFM
and Q$_M$ AFM with the optimized structure of LaOFeSb at several volumes from FPLO7 with PW92 XC functional. Upper part: z(Sb) is
 optimized. Lower part: z(Sb) is optimized and shifted.
}
\label{Sb-mag-th}
\end{table}

At 1.138 V$_0$, the inferred equilibrium volume of LaOFeSb, the properties of NM/FM,
Q$_0$ AFM, and Q$_M$ AFM are very similar to the ones of LaOFeAs at its experimental volume.
Thus from these results we expect that doped LaOFeSb should have similar properties (viz, value of T$_c$) as LaOFeAs.

\subsection{LaOFeN}
The structure of LaOFeN is also not reported experimentally. In order to obtain it,
the same procedure as for LaOFeSb was used. The lowest total energy is at 0.762 V$_0$' 
(here V$_0$' is the experimental volume of LaOFeP.). Again assuming PW92 makes a similar error
as it makes in LaOFeAs, we estimate its equilibrium 
volume to be close to 0.825 V$_0$'. At 0.825 V$_0$' and for larger volume, the total energy of the 
Q$_M$ AFM state is well below that of the FM/NM state (see Table \ref{N-mag-ex}). Therefore,
LaOFeN, if it exists, should be in the Q$_M$ AFM ordered state at low temperature, which is similar to
LaOFeAs and LaOFeSb.

\begin{table}
\begin{tabular}{|c|c|c|c|c|c|}
\hline
V/V$_0$' & \multicolumn{3}{|c|}{mag. mom. ($\mu$$_B$)} & \multicolumn{2}{|c|}{$\Delta$ EE (meV/Fe)} \\
\hline
             & Q$_M$ & Q$_0$ & NM/FM  & NM/FM-Q$_M$ & Q$_0$-Q$_M$ \\
0.900 &	2.21 & 1.69 & 1.64 & 209.8 & 135.9 \\ 
0.875 &	2.06 & 1.51 & 0.03 & 114.9 & 99.2  \\
0.850 &	1.88 & 1.14 & 0.03 & 74.3  & 68.1  \\
0.825 &	1.63 & 0.80 & 0.03 & 41.0  & 40.0  \\
0.800 &	1.26 & --- & 0.00 & 18.4  & ---  \\
0.787 &	1.08 & --- & 0.00 & 11.3  & ---  \\
0.775 &	0.90 & --- & 0.00 & 7.0   & ---  \\
0.762 &	0.00 & --- & 0.00 & 1.3   & ---  \\
0.750 &	0.00 & --- & 0.00 & 1.4   & ---  \\
0.725 & 0.00 & --- & 0.00 & 1.2   & ---  \\
0.700 &	0.00 & --- & 0.00 & 0.9   & ---  \\
\hline
\end{tabular}
\caption{Calculated magnetic moment of Fe in LaOFeN, total energy relative to the nonmagnetic (ferromagnetic) states of Q$_0$ AFM
and Q$_M$ AFM with the optimized structure at several volumes, but shifted z(N) up by 0.011,
as a compensation PW92 does to LaOFeAs, where PW92 underestimates z(As) by 0.011.}
\label{N-mag-ex}
\end{table}

\begin{figure}[tbp]
\rotatebox{-90}
{\resizebox{6.0cm}{8.6cm}{\includegraphics{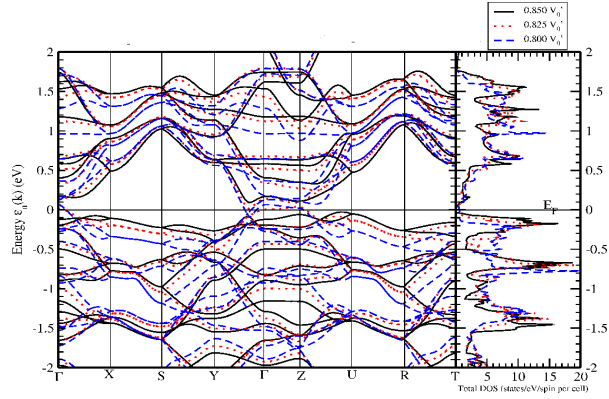}}}
\caption{ Plot of LaOFeN Q$_M$ AFM band structure and total DOS
 at 0.850V$_0$', 0.825 V$_0$' and  0.800 V$_0$' with shifted z(N).}
\label{N-band-DOS-QM-P}
\end{figure}

Compared to the other LaOFePn compounds, LaOFeN is even closer to being a semimetal when the volume is equal to 0.825 V$_0$',
and it becomes a small gap insulator at 0.850 V$_0$' and a higher carrier density metal at 0.800 V$_0$' (see Fig. \ref{N-band-DOS-QM-P}).
The DOS for 0.825 V$_0$' shows a pseudogap around E$_F$, but the DOS is somewhat less flat than it is
for LaOFeAs.

When LaOFeN is calculated to be insulating (for volumes larger than 0.825 V$_0$'), the gap can be taken to {\it define} a distinction
between bonding (occupied) and antibonding (unoccupied) states. 
The appearance of this gap in LaOFeN is quite surprising: although there is clear separation of valence 
and conduction bands over most of the zones for LaOFeAs, there is no way to ascribe the small FSs to simple
overlapping valence and conduction bands: in LaOFeAs and LaOFeSb, the bonding and antibonding bands are never completely 
  separated from each other. In LaOFeN this separation finally becomes apparent, as an actual bandgap does appear.

\section{Role of the rare earth atom in ReOFeAs}
After LaOFeAs was discovered, after appropriate variation of the carrier concentration,
to be superconducting at 26 K, much substitution on the rare earth (R) site has
been done, with impressive increases in the critical temperature.  Since all are
evidently trivalent and donate three valence electrons to the FeAs layer, it becomes
important to uncover the influence of the R atom: is it some aspect
of the chemistry, which does differ among the rare earths? is it an effect of
size? or can there be some other subtle effect?

  Table \ref{Re-Tc} is a collection of the lattice constants $a$ and $c$, volume V of the primitive cell,
T$_c$ onset
of ROFeAs reported from experiment.\cite{ReOFeAs-Tc1, ReOFeAs-Tc2, ReOFeAs-Tc3, ReOFeAs-Tc4} 
Both lattice constants, hence the volume, 
decrease monotonically as the atomic number increases, but T$_c$ increases only from La to Gd, whereupon drops for heavier rare earths. 
Since we have found that small details affect the electronic and magnetic structure -- especially $z$(As) --
it is reasonable to assess the size effect. We have performed calculations 
on Ce, Nd and Gd, using LSDA+U with U=7.0 eV and J=1 eV applied to the R atom to occupy the $4f$ shell
appropriately and keep the $4f$ states away from the Fermi level. Our results indicate that
all have very similar DOS and band structure with LaOFeAs. To investigate further, we checked GdOFeAs using 
the crystal structure of LaOFeAs. The resulting band structure and DOS 
are almost identical to the original results for Gd, thus there seems to be no appreciable effect of the
differing chemistries of Gd and La. This negative result supports the idea that the size difference may be dominant,
though seemingly small. 
The difference in size (hence $a$, $c$, and the internal coordinates) influences not only the 
band structure and DOS, but also the magnetic properties. 
Fixed spin moment calculations in the FM state gives the lowest total energy at 0.2 $\mu_B$/Fe in LaOFeAs, 
and 0.5 $\mu_B$/Fe in both GdOFeAs and La-replaced GdOFeAs.

\begin{table}
\caption{Collection of the lattice constants a (\AA) and c(\AA), 
volume V (\AA$^3$ of the primitive cell,
T$_c$ onset
s (onset, middle, and zero, in K) 
of ReOFeAs reported from experiments. }
\label{Re-Tc}
\begin{tabular}{|c|c|c|c|c|c|}
\hline
element & Z   & a(\AA) & c(\AA) & V(\AA$^3$) & T$_{C,onset}$ (K)\\ 
La      & 57  & 4.033  & 8.739  & 142.14     & 31.2           \\
Ce      & 58  & 3.998  & 8.652  & 138.29     & 46.5           \\
Pr      & 59  & 3.985  & 8.595  & 136.49     & 51.3           \\
Nd      & 60  & 3.965  & 8.572  & 134.76     & 53.5           \\
Sm      & 62  & 3.933  & 8.495  & 131.40     & 55.0           \\
Gd      & 64  & 3.915  & 8.447  & 129.47     & 56.3           \\
Tb      & 65  & 3.899  & 8.403  & 127.74     & 52             \\
Dy      & 66  & 3.843  & 8.284  & 122.30     & 45.3           \\
\hline
\end{tabular}
\end{table}

\section{Summary}

We have investigated in some detail the electronic structure and magnetic properties of the LaOFeAs 
class of novel superconductors using ab-initio methods.
 The effects of the Fe-As distance, of doping, and of pressure, as well as calculations of the EFGs have been reported.
  It was found that (approximate) electron-hole symmetry versus doping, and strong magnetophonon coupling 
   are primary characteristics 
   of the LaOFeAs system, and are two of the ingredients that need to be understood to proceed toward the discovery
    the mechanism of superconducting pairing.
     We studied effects of the structural distortion and of the ($\pi,\pi,\pi$) magnetic order,
     finding that experiments can be reproduced fairly well by our calculations. Finally, the related materials LaOFeP, LaOFeSb, and
      LaOFeN were investigated and their properties compared to those of LaOFeAs. From these comparisons, it appears that LaOFeP is significantly 
       different from the other materials studied here; this difference might explain why, at stoichiometry, 
        LaOFeP is superconducting while LaOFeAs is antiferromagnetic.
       Also, in view of their similarities with LaOFeAs, either pure or doped LaOFeSb and LaOFeN are potential candidates as superconductors.

\section{Acknowledgments}

We acknowledge financial support from ANR PNANO grant N$^{\text{O}}$
  ANR-06-NANO-053-02 and N$^{\text{O}}$ ANR-BLAN07-1-186138 (S.L.) and
from DOE Grant DE-FG03-01ER45876 (W.E.P.). W.E.P. is happy to acknowledge a grant from the France
Berkeley Fund that enabled the initiation of this project.

\end{document}